\documentclass[twocolumn,showpacs,pre]{revtex4}
\usepackage{amssymb}
\usepackage{makeidx}
\usepackage{amsmath}
\usepackage{graphicx}
\usepackage{dcolumn}
\usepackage{bm}
\usepackage{epstopdf}

\begin{document}

\title{Ultrarelativistic electron bunches of solid densities and nuclear
radiation from nanolayers-plasma-targets under superintense laser pulses }
\author{H.K. Avetissian$^{1}$}
\author{H.H. Matevosyan$^{2}$}
\author{G.F. Mkrtchian$^{1}$}
\author{Kh.V. Sedrakian$^{1}$}
\affiliation{$^{1}$ Centre of Strong Fields Physics, Yerevan State University, 1 A.
Manukian, Yerevan 0025, Armenia}
\affiliation{$^{2}$ Plasma Theory Group, Institute of Radiophysics and Electronics, 0203
Ashtarak, Armenia}
\date{\today }

\begin{abstract}
We consider nonlinear interaction of superpower laser pulses of relativistic
intensities with nanolayers and solid-plasma-targets towards the production
of high energy-density electron bunches along with nuclear radiation (hard $%
\gamma $-quanta and positron fluxes). It is shown that petawatt lasers are
capable of producing via two-target scheme high density field-free
electron/positron bunches and substantial amounts of $\gamma $-quanta with
energies up to $200$ MeV. For actual supershort and tightly
focused--strongly nonplane ultrarelativistic laser pulses of linear and
circular polarizations 3D3V problem is solved via numerical simulations.
\end{abstract}

\date{\today }
\maketitle

\section{Introduction}

Acceleration of electrons with superpower laser beams (at present --of
relativistic intensities)\ and their interaction with the matter in
ultrashort space-time scales have attracted broad interest over the last two
decades, stimulated by the continued progress made in laser technology. In
many laboratories \cite{Lab1,Lab2,Lab3,Lab4} compact lasers can deliver $%
1-10 $ petawatt short pulses, with focused intensities as high as $10^{22}\ 
\mathrm{W\ cm}^{-2}$. Next generation multipetawatt and exawatt optical
laser systems \cite{ELI,XCELS,HiPER} will be capable to deliver ultrahigh
intensities exceeding $10^{25}\ \mathrm{W\ cm}^{-2}$, which are well above
the ultrarelativistic regime of electron-laser interaction. Laser
intensities at which an electron becomes relativistic is defined by the
condition $\xi \gtrsim 1$, where $\xi =eE\lambdabar /mc^{2}$ is the
relativistic invariant dimensionless parameter of a wave-particle
interaction ($e$ is the elementary charge, $c$ is the light speed in vacuum,
and $m$ is the electron mass) and represents the work of the wave electric
field with strength $E$ on a wavelength $\lambda $ ($\lambdabar =\lambda
/2\pi $) in the units of electron rest energy. Laser intensities expressed
via $\xi $ can be estimated as:%
\begin{equation*}
I_{r}=\xi ^{2}\times 1.37\times 10^{18}\ \mathrm{W\ cm}^{-2}[\lambda /%
\mathrm{\mu m}]^{-2}.
\end{equation*}%
Thus, available optical lasers provide $\xi $ up to $100$, meanwhile with
the next generation of laser systems one can manipulate with beams at $\xi $ 
$>1000$. In such ultrastrong laser fields, electrons can reach to
ultrarelativistic energies. However to obtain ultrarelativistic net energies
for field-free electrons one should bypass the limitations imposed by the
well known Lawson-Woodward theorem \cite{LW}. One way to get net energy
exchange is to use coherent processes of laser-particle interaction with the
additional resonances. Among those the induced Cherenkov, Compton, and
undulator processes are especially of interest \cite{Book,accel}. In these
induced processes one can obtain moderate relativistic electron beams of low
energy spreads and emittances due to the threshold character of nonlinear
resonance in the strong wave field. Besides, the use of a plasma medium \cite%
{Plasma} is a promising way of achieving laser-driven electron acceleration.
However, laser-plasma accelerator schemes face problems connected with the
inherent instabilities in laser-plasma interaction processes. The spectrum
of direct acceleration mechanisms of charged particles by a single laser
pulse is very restricted, since one should use the laser beams focused to
subwavelength waist radii, or use subcycle laser pulses, or use radially
polarized lasers. All these scenarios with different field configurations
have been investigated both theoretically and experimentally \cite%
{A1,A2,A3,A4,A5,A6,A7,A8,A9,A10,A11,A12,A13,A14,A15,A16,A17}. The
Lawson-Woodward theorem can also be bypassed by terminating the field,
either by reflection, absorption, or diffraction \cite{Pantell}. The proof
of principle experiment of this type has been reported in Ref. \cite{PPP}.
Here it is used initially relativistic electron beam and a moderately strong
laser pulse. To obtain dense enough electron bunches it is reasonable to
consider electrons acceleration from nanoscale-solid plasma-targets \cite%
{mirror1,mirror2,mirror3,mirror4} at intensities high enough to separate all
electrons from ions \cite{Kulagin}. Thus, combining these two schemes one
can obtain ultrarelativistic solid density electron bunches. Such bunches
can be used to obtain high-flux of positrons, $\gamma $-quanta with possible
applications in material science, medicine, and nuclear physics.

In the present work we propose an efficient mechanism for creation of
ultrarelativistic field-free electron bunches, gamma-ray and positron beam
of high fluxes by a single laser pulse of ultrarelativistic intensities. The
scheme is as follows: a laser beam of ultrarelativistic intensities is
focused onto nanoscale-solid plasma-target with relatively low $\mathcal{Z}$%
. From this target under the action of ultrashort laser pulse a superdense
electron bunch is formed and accelerated up to ultrarelativistic energies.
Then we place a high-$\mathcal{Z}$ target (tungsten or gold) at the
distances where the electron bunch gains maximum energy from the laser
pulse. The purpose of the second target is twofold. First, it acts as a
reflector for laser pulse, being practically transparent for electron bunch.
As a consequence we get field-free ultrarelativistic electron bunch. Second,
within this target with sufficiently large thickness the generation of hard $%
\gamma $-quanta and positron fluxes takes place due to the electron-ion
collisions.

The organization of the paper is as follows. In Sec. II for supershort
strongly nonplane ultrarelativistic laser pulses of linear and circular
polarizations 3D3V problem is solved via numerical simulations. In Sec. III
we present numerical calculations for produced $\gamma $-quanta and positron
fluxes. Finally, conclusions are given in Sec. IV.

\section{Generation of ultrarelativistic electron bunches from nanotargets}

Here we report on the results of the 3D3V simulations of superintense laser
beam interaction with solid-plasma-targets. The illustration of proposed
scheme is as follows: a laser beam of ultrarelativistic intensities ($\xi
>>1 $) is focused onto solid-plasma-target of nanothickness with relatively
low $\mathcal{Z}$. For concreteness we take carbon target ($\mathcal{Z}=6$).
Target is assumed to be fully ionized. This is justified since the intensity
for full ionization of carbon is about $10^{19}$ $\mathrm{W/cm}^{2}$, while
we use in simulations intensities at least on two order of magnitude larger.
So, the target will become fully ionized before the arrival of the pulse
peak. Electrons are assumed to be cold in the target, $T_{e}=0$. The carbon
target size is $50\lambda \times 50\lambda $ in the $xy$ plane and $%
d=0.002\lambda $ in $z$ direction (wave propagation direction). The foil is
chosen to be thin enough for the laser beam to push out all electrons from
the target. In this case, only electrons are accelerated, while the ions are
left unmoved. For this regime \cite{Kulagin}, the relativistic invariant
parameter of the laser field $\xi $ must be larger than the normalized field
arising from electron-ion separation: $\eta =2\pi \left( n_{e}/n_{c}\right)
d/\lambda $. Here $n_{e}$ is the electron density ($n_{e}\simeq 6.5\times
10^{23}\mathrm{cm}^{-3}$ for carbon) and $n_{c}=m\omega ^{2}/(4\pi e^{2})$
is the critical density. We assume laser radiation of $\lambda =800\ \mathrm{%
nm}$ and therefore $n_{c}\simeq 1.74\times 10^{21}\ \mathrm{cm}^{-3}$. For
the chosen thickness $d=\allowbreak 1.\,\allowbreak 6\ \mathrm{nm}$ we have $%
\eta \simeq 4.7$. Thus, to fulfill the condition $\xi >$ $\eta $ for all
calculations we take $\xi =20$ for a laser beam of circular polarization and 
$\xi =30$ -for a linear one. Although the electron density is overcritical,
the target is much thinner than the skin depth and is therefore transparent
for the laser beam. Thus, from this target under the action of ultrashort
laser pulse a superdense electron bunch as a thin sheath is formed and
accelerated up to ultrarelativistic energies. Hence, their motion is well
described in a single-particle picture, i.e. we will solve relativistic
equations of motion for macroparticles 
\begin{equation}
\frac{d\mathbf{\Pi }}{dt}=\frac{e}{mc}\left( \mathbf{E}+\frac{\mathbf{\Pi }%
\times \mathbf{H}}{\gamma }\right) ,\ \frac{d\gamma }{dt}=\frac{e}{mc}\frac{%
\mathbf{\Pi }\cdot \mathbf{E}}{\gamma }  \label{1}
\end{equation}%
in the given electric $\mathbf{E}$ and magnetic $\mathbf{H}$ fields. Here we
have introduced normalized momentum $\mathbf{\Pi }\ =\mathbf{p/(}mc)$,
energy $\gamma =\sqrt{1+\mathbf{\Pi }^{2}}$ (Lorentz factor) and have taken
into account that charge to mass ratio of macroparticles is the same as for
the electron.

The laser beam is focused at the $z=0$ and propagates along the $OZ$. For
analytic description of such pulses of linear/circular polarization we will
approximate corresponding electromagnetic fields as follow\cite{Mcdon}:%
\begin{equation}
E_{x}=\widetilde{E}_{0}\left( \mathbf{r},t\right) \left( \cos \varphi _{-}-%
\frac{z}{z_{R}}\sin \varphi _{-}\right) ,  \label{Ex}
\end{equation}%
\begin{equation}
E_{y}=g\widetilde{E}_{0}\left( \mathbf{r},t\right) \left( \sin \varphi _{-}+%
\frac{z}{z_{R}}\cos \varphi _{-}\right) ,  \label{Ey}
\end{equation}%
\begin{equation*}
E_{z}=\frac{\widetilde{E}_{0}\left( \mathbf{r},t\right) \lambda }{\pi
w^{2}\left( z\right) }\left[ \left( gy\left( 1-\frac{z^{2}}{z_{R}^{2}}%
\right) -2\frac{zx}{z_{R}}\right) \cos \varphi _{-}\right.
\end{equation*}%
\begin{equation}
\left. +\left( x\left( 1-\frac{z^{2}}{z_{R}^{2}}\right) +2g\frac{zy}{z_{R}}%
\right) \sin \varphi _{-}\right] ,  \label{Ez}
\end{equation}%
\begin{equation}
H_{x}=-E_{y},\qquad H_{y}=E_{x},  \label{Hxy}
\end{equation}%
\begin{equation*}
H_{z}=\frac{\widetilde{E}_{0}\left( \mathbf{r},t\right) \lambda }{\pi
w^{2}\left( z\right) }\left[ \left( -gx\left( 1-\frac{z^{2}}{z_{R}^{2}}%
\right) -2\frac{zy}{z_{R}}\right) \cos \varphi _{-}\right.
\end{equation*}%
\begin{equation}
\left. +\left( y\left( 1-\frac{z^{2}}{z_{R}^{2}}\right) -2g\frac{zx}{z_{R}}%
\right) \sin \varphi _{-}\right] ,  \label{Hz}
\end{equation}%
where 
\begin{equation}
\widetilde{E}_{0}\left( \mathbf{r},t\right) =E_{0}\frac{w_{0}^{2}}{%
w^{2}\left( z\right) }e^{-\frac{\rho ^{2}}{w^{2}\left( z\right) }}f\left(
t-z/c\right)  \label{ampl}
\end{equation}%
is the envelope function, $g$ is the parameter of ellipticity; $g=0$
corresponds to a linear polarization and $g=1$ -- circular polarization, $%
E_{0}$ is the electric field amplitude, $\rho ^{2}=x^{2}+y^{2}$, $w\left(
z\right) =w_{0}\sqrt{1+z^{2}/z{}_{R}^{2}}$, where $z_{R}=\pi
w_{0}^{2}/\lambda $ is the Rayleigh length of the focused laser pulse with
the waist $w_{0}$ in the focal plane $z=0$, and $\varphi _{-}=\omega
t-kz\left( 1+w_{0}^{2}\rho ^{2}/\left( 2z^{2}w^{2}\left( z\right) \right)
\right) $ is the deformed phase. The laser pulse has temporal profile $%
f\left( t\right) =\cosh \left( t/\mathcal{T}\right) $ with the pulse
durations $\mathcal{T}=4\lambda /c$. 
\begin{figure}[tbp]
\includegraphics[width=.5\textwidth]{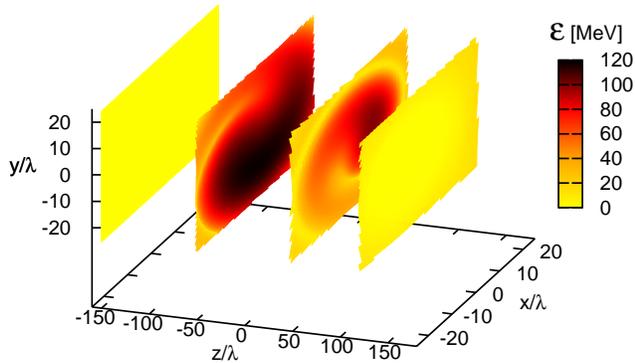}
\caption{(Color online) Energy distribution in accelerated electron layer is
shown at various $z$ for circularly polarized wave with $w_{0}=40\protect%
\lambda $ and $\protect\xi =20$. There is no reflector-target.}
\label{eps1}
\end{figure}
The carbon target is situated at $z_{t}=-150\lambda $. For the numerical
calculations the total number of macroparticles is taken to be $4\times
10^{4}$ and uniformly distributed in the target. The set of equations for
the macroparticles has been solved using a standard fourth-order
Runge--Kutta algorithm.

As the employed laser pulse is of nonplane configuration, here there is
acceleration effect of particles even in the field of a single pulse.
However, the latter is small in accordance with the above mentioned
Lawson-Woodward theorem \cite{LW}, to assess of which we made simulations
with a single pulse given by Eqs. (\ref{Ex})-(\ref{Hz}). For this setup the
energy distribution in accelerated electron layer is shown in Fig. 1 for
various $z$. Calculation have been made for the circularly polarized wave
with $w_{0}=40\lambda $ and $\xi =20$ ($I_{\max }\simeq 1.7\times 10^{21}\ 
\mathrm{W\ cm}^{-2}$). As Fig. 1 evidences, the residual acceleration
because of nonplane character of a focused laser beam is small. Meantime, as
is seen from this figure in the laser field electrons reach
ultrarelativistic energies. Thus, to keep the energy gained from the wave
field we propose to place a high-$\mathcal{Z}$ target near the position $%
z=z_{m}$ where electrons gain maximal amount of energy from the laser pulse.
The purpose of the second target is twofold. First, it acts as a reflector
for laser pulse, being practically transparent for electron bunch. As a
consequence, we will get field-free ultrarelativistic electron bunch in the
high-$\mathcal{Z}$ target. Second, within this target with sufficiently
large thickness the generation of $\gamma $-ray and positrons take place due
to the electron bunch--high-$\mathcal{Z}$ ion collisions. Due to high
density and thickness (which should be much larger than skin depth) of
second target we will consider it as a perfect reflector. Hence, one should
solve the equations (\ref{1}) taking into account also reflected wave. For
the total electric $\mathbf{E}^{(\mathrm{tot})}$ and magnetic $\mathbf{H}^{(%
\mathrm{tot})}$ fields we will have: for\textrm{\ }$z>z_{m}$%
\begin{equation}
\mathrm{\ \ }\mathbf{E}^{(\mathrm{tot})}=0,\ \mathbf{H}^{(\mathrm{tot})}=0
\label{after}
\end{equation}%
while for $z\leq z_{m}$%
\begin{equation}
E_{x,y}^{(\mathrm{tot})}\left( \mathbf{r},t\right) =E_{x,y}\left(
x,y,z,t\right) -E_{x,y}\left( x,y,-z+2z_{m},t\right) ,  \label{Exyt}
\end{equation}%
\begin{equation}
E_{z}^{(\mathrm{tot})}\left( \mathbf{r},t\right) =E_{z}\left( x,y,z,t\right)
+E_{z}\left( x,y,-z+2z_{m},t\right) ,  \label{Ezt}
\end{equation}%
\begin{equation}
H_{x,y}^{(\mathrm{tot})}\left( \mathbf{r},t\right) =H_{x,y}\left(
x,y,z,t\right) +H_{x,y}\left( x,y,-z+2z_{m},t\right) ,  \label{Hxyt}
\end{equation}%
\begin{equation}
H_{z}^{(\mathrm{tot})}\left( \mathbf{r},t\right) =H_{z}\left( x,y,z,t\right)
-H_{z}\left( x,y,-z+2z_{m},t\right) .  \label{Hzt}
\end{equation}%
The typical picture of acceleration dynamics is shown in Fig. 2, where
energy distribution in accelerated electron layer is shown for various $z$
for the circularly polarized wave with $w_{0}=40\lambda $. The
reflector-target is situated at $z_{m}=0$. As is seen from Fig. 2, after the
passing the second target we have field-free ultrarelativistic electron
bunch. 
\begin{figure}[tbp]
\includegraphics[width=.5\textwidth]{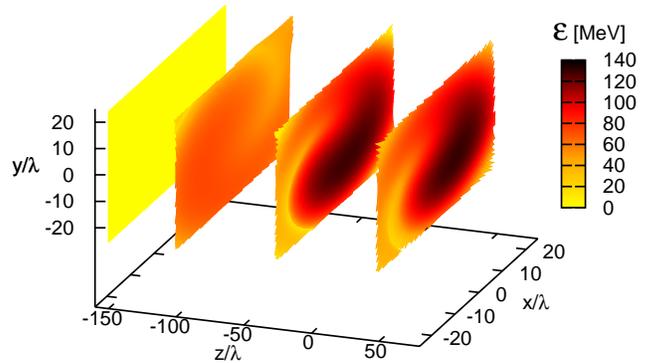}
\caption{(Color online) Energy distribution in accelerated electron layer is
shown at various $z$ for circularly polarized wave with $w_{0}=40\protect%
\lambda $ and $\protect\xi =20$. Reflector-target is situated at $z_{m}=0$.}
\end{figure}
\begin{figure}[tbp]
\includegraphics[width=.5\textwidth]{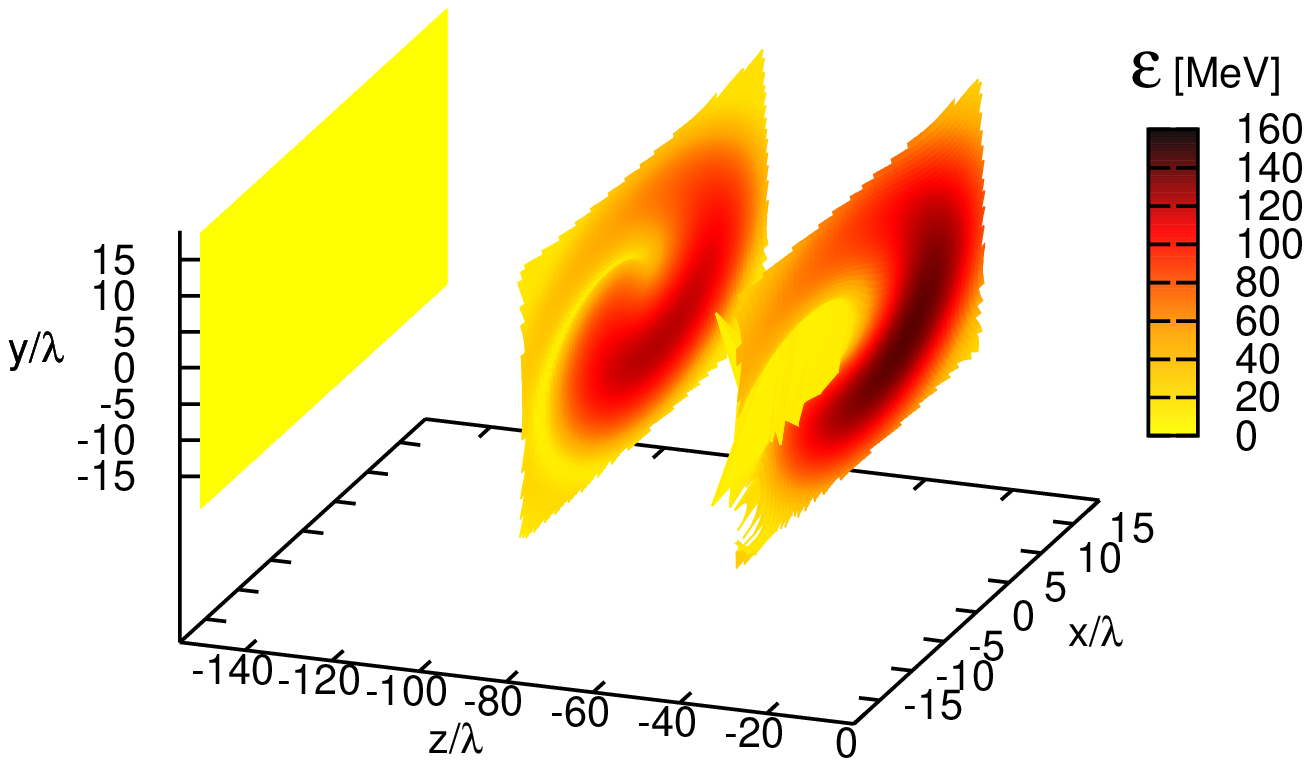}
\caption{(Color online) Energy distribution in accelerated electron layer at
various $z$ for circularly polarized wave with $w_{0}=15\protect\lambda $
and $\protect\xi =20$. Reflector-target is situated at $z_{m}=-50\protect%
\lambda $.}
\end{figure}
\begin{figure}[tbp]
\includegraphics[width=.5\textwidth]{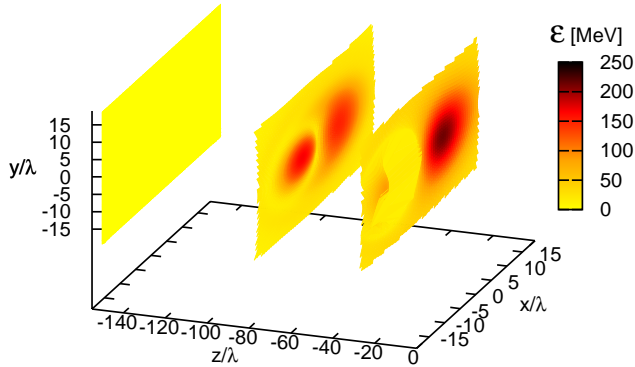}
\caption{(Color online) Energy distribution in accelerated electron layer at
various $z$ for linearly polarized wave with $w_{0}=10\protect\lambda $ and $%
\protect\xi =30$. Reflector-target is situated at $z_{m}=-50\protect\lambda $%
.}
\end{figure}

For the smaller waists $w_{0}$ the position of energy maximum takes place
for the smaller interaction lengths $\left\vert z_{t}-z_{m}\right\vert $. In
Fig. 3, we represent the results for circularly polarized wave with $%
w_{0}=15\lambda $ and $\xi =20$. For this setup the reflector-target is
situated at $z_{m}=-50\lambda $. We have made calculations also for the
linearly polarized laser beam. In Fig. 4, the results for linearly polarized
laser beam ($g=0$) with $w_{0}=10\lambda $ and $\xi =30$ ($I_{\max }\simeq
1.9\times 10^{21}\ \mathrm{W\ cm}^{-2}$) are presented. As is seen from this
figure, in the field-free region we have an ultrarelativistic electron bunch
of energies $\sim 200\ \mathrm{MeV}$.

Note that in the high-$\mathcal{Z}$ target due to the electron bunch--ion
collisions the electrons will lose energy via the generation of $\gamma $%
-quanta and positrons. Hence, if one expects to obtain unperturbed electron
bunch, the target thickness should be much smaller than the radiation
length, otherwise along with primary bunch one will have substantial amounts
of produced gamma-ray photons and positrons. This scenario is also of
interest and will be considered in the next section.

\section{Production of hard $\protect\gamma $-quanta and positrons with
laser generated ultrarelativistic electron bunches}

As is seen from the figures 1-4, the energy distribution of accelerated
electrons obtained by the laser-target interaction is usually broad with
almost $100\%$ energy spread up to a cutoff energy. Such electron bunches
are not suitable for coherent effects, for example, to generate coherent
radiation -free electron laser. However, such bunches of high energy-density
can be used to obtain high-flux of positrons, hard $\gamma $-quanta with
possible applications in material science, medicine, and nuclear physics.
Note that in our scheme this can be implemented by all-optical means without
use of large scale facilities such as nuclear reactors or conventional
particle accelerators. One can use second target as a bremsstrahlung
converter, leading to the emission of bremsstrahlung photons, the energy of
which takes a maximum value equal to that of the accelerated electrons.
Then, the bremsstrahlung photons interacting with ions can produce
electron-positron pairs. In other words, QED cascade can be developed.
Accurate calculations of $\gamma $-quanta and positrons spectra in such
targets require using Monte Carlo computer codes (see, e.g., Ref. \cite%
{Monte} and references therein). However, one can use relatively simple
formulas \cite{Rossi} for thin targets, when target thickness is smaller
than the radiation length $\mathcal{L}_{\mathrm{rad}}$:%
\begin{equation}
\mathcal{L}_{\mathrm{rad}}=\left( 4\alpha N_{i}\mathcal{Z}^{2}r_{0}^{2}\log 
\frac{183}{\mathcal{Z}^{1/3}}\right) ^{-1},  \label{RadL}
\end{equation}%
where $\alpha =e^{2}/\hbar c=1/137$ is the fine structure constant, $N_{i}$
is the density of ions, $r_{0}$ - is the electron classical radius. In
particular, for gold ($\mathcal{Z}=79$, $N_{i}=5.\,\allowbreak 9\times
10^{22}\ \mathrm{cm}^{-3}$) $\mathcal{L}_{\mathrm{rad}}\simeq 0.31\ \mathrm{%
cm}$, and for tantalum ($\mathcal{Z}=73$, $N_{i}=5.\,\allowbreak 55\times
10^{22}\ \mathrm{cm}^{-3}$) $\mathcal{L}_{\mathrm{rad}}\simeq 0.39\ \mathrm{%
cm}$\textrm{. }Thus, the $\gamma $-quanta number distribution $N_{\gamma }$
over energy $\varepsilon _{\gamma }$ obeys integro-differential equation 
\cite{Rossi}:%
\begin{equation}
\frac{\partial N_{\gamma }\left( \varepsilon _{\gamma },\zeta \right) }{%
\partial \zeta }=\int\limits_{0}^{1}N_{e}\left( \frac{\varepsilon _{\gamma }%
}{\upsilon },\zeta \right) \varphi _{0}\left( \upsilon \right) \frac{%
d\upsilon }{\upsilon }-\sigma _{0}N_{\gamma }\left( \varepsilon _{\gamma
},\zeta \right) ,  \label{gamma}
\end{equation}%
Here $\sigma _{0}\simeq 7/9$, and $\varphi _{0}\left( \upsilon \right) $
describes bremsstrahlung process, which in the case of complete screening is
given by the formula:%
\begin{equation}
\varphi _{0}\left( \upsilon \right) =\frac{1}{\upsilon }\left( 1+\left(
1-\upsilon \right) ^{2}-\varkappa _{0}\left( 1-\upsilon \right) \right) ,
\label{fi}
\end{equation}%
where $\varkappa _{0}\simeq 0.64$. The quantity $\zeta $ in Eq. (\ref{gamma}%
) is the target thickness measured in radiation lengths, $N_{e}\left( 
\mathcal{E},\zeta \right) $ is the electron/positron distribution function,
that should be defined self-consistently. However, for small $\zeta $, in
Eq. (\ref{gamma}) we take $N_{e}\left( \mathcal{E},\zeta \right)
=N_{e}\left( \mathcal{E},0\right) $, where $N_{e}\left( \mathcal{E},0\right) 
$ is the distribution function at the entrance to high-$\mathcal{Z}$ target.
Thus, for small $\zeta $ we have 
\begin{equation}
N_{\gamma }\left( \varepsilon _{\gamma },\zeta \right) \simeq \zeta
\int\limits_{0}^{1}N_{e}\left( \frac{\varepsilon _{\gamma }}{\upsilon }%
,0\right) \varphi _{0}\left( \upsilon \right) \frac{d\upsilon }{\upsilon }.
\label{gam}
\end{equation}%
\qquad \qquad \qquad In the limit of low annihilation rates the positrons
distribution function can be defined as 
\begin{equation}
\frac{\partial N_{e^{+}}\left( \mathcal{E},\zeta \right) }{\partial \zeta }%
=\int\limits_{0}^{1}N_{\gamma }\left( \frac{\mathcal{E}}{u},\zeta \right)
\psi _{0}\left( u\right) \frac{du}{u},  \label{pos}
\end{equation}%
where $\psi _{0}\left( u\right) $ describe pair production process, which in
the case of complete screening is given by the formula \cite{Rossi}: 
\begin{equation}
\psi _{0}\left( u\right) =u^{2}+\left( 1-u\right) ^{2}+\varkappa _{0}u\left(
1-u\right) .  \label{psi}
\end{equation}%
\begin{figure}[tbp]
\includegraphics[width=.43\textwidth]{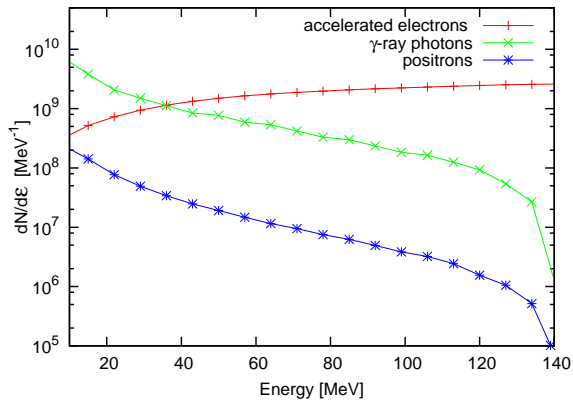}
\caption{(Color online) In the logarithmic scale it is shown the energy
distribution functions for primary accelerated electrons, produced $\protect%
\gamma $-quanta and positrons for the setup of Fig. 3. }
\end{figure}
\begin{figure}[tbp]
\includegraphics[width=.43\textwidth]{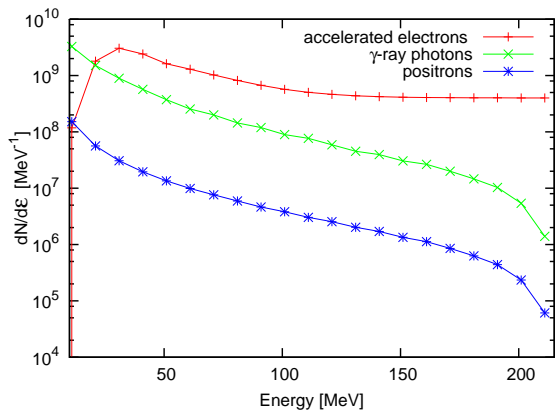}
\caption{(Color online) Same as Fig. 5, but for the setup of Fig. 4. }
\end{figure}

Thus, having at hand electron bunch distribution $N_{e}\left( \mathcal{E}%
,0\right) $ with the help of Eqs. (\ref{gam}) and (\ref{pos}) one can
calculate hard $\gamma $-quanta and positrons number distribution over
energy. For the target thickness we take $\zeta =0.2$. We have made
calculations for the setup of figures 3 and 4. To obtain electron bunch
distribution $N_{e}\left( \mathcal{E},0\right) $ at the entrance to
bremsstrahlung target we considered the effective area limited to the
central zone between $-20\lambda <x,y<20\lambda $. The results of
calculations are presented in Figs 5 and 6. In these figures, in the
logarithmic scale it is shown the energy distribution functions for primary
accelerated electrons, produced $\gamma $-quanta, and positrons. As is seen
from these figures, through two-target scheme one can produce dense
electron/positron bunches and substantial amounts of $\gamma $-quanta of
energies up to $200$ MeV. In particular, for Fig. 6 the number of photons in
the range of $50-200\ \mathrm{Mev}$ is $\sim 10^{9}$.

\section{Conclusion}

We have proposed mechanism for generation of high energy-density particles
bunches from nanotargets by single laser pulse of ultrarelativistic
intensities. We consider two target scheme where one ultrathin target serve
as a source of dense electron bunch. The purpose of the second target is
twofold. First, it acts as a reflector for laser pulse, thus abruptly
terminating the wave field and therefore allows electrons to keep the gained
from the wave energy. Second, within this target with sufficiently large
thickness the intensive generation of hard $\gamma $-quanta and positrons
occurs due to the electron-ion collisions. In particular, the considered
setup provides generation of high energy-density positrons and $\gamma $%
-quanta that may have applications in material science, medicine, and
nuclear physics.

\begin{acknowledgments}
This work was supported by SCS of RA.
\end{acknowledgments}

\end{document}